\documentclass[traditabstract,twocolumn]{aa}

\usepackage{graphicx}
\usepackage{txfonts}
\usepackage{natbib}

\newcommand{\rev}[1]{{ #1}}
\newcommand\micron{\mbox{$\mu$m}}%

\begin{document}

\title{
The Dust Mantle of Comet 9P/Tempel 1:
Dynamical Constraints on Physical Properties 
}

\subtitle{}



\author{H. Kobayashi
\inst{1}
\and
H. Kimura
\inst{2}
\and
S. Yamamoto
\inst{3}
}

\institute{
Department of Physics, Nagoya University, Nagoya, Aichi464-8602, Japan\\
\email{hkobayas@nagoya-u.jp}
\and 
Center for Planetary Science, c/o Integrated Research Center of Kobe
University, Chuo-ku Minatojima Minamimachi 7-1-48, Kobe 650-0047, Japan\\
\and
Center for Environmental Measurement and Analysis, National Institute
for Environmental Studies, 16-2 Onogawa, Tsukuba, Ibaraki 305-8506,
Japan\\
}

\date{}


\abstract{
The trajectories of dust particles ejected from a comet are affected by
solar radiation pressure as a function of their ratios of radiation
pressure cross section to mass. Therefore, a study on the orbital
evolution of the particles caused by the radiation pressure reveals the
physical properties of dust on the surface of the comet nucleus. In the
course of NASA's Deep Impact mission, the ejecta plume evolved under the
influence of the radiation pressure. From the evolution and shape of
the plume, we have succeeded in obtaining $\beta \approx 0.4$, where
$\beta$ is the ratio of the radiation pressure to the solar gravity.
Taking into account $\beta \approx 0.4$ as well as the observational
constraints of a high color temperature and a small silicate-feature
strength, dust particles ejected from the surface of comet 9P/Tempel 1
are likely compact dust aggregates of sizes $\approx 20\,\mu$m  (mass
 $\sim 10^{-8}$\,g).  This is comparable to the major dust on the
 surface of comet 1P/Halley ($\sim 10\mu$m) inferred from in-situ
 measurements and theoretical considerations. Since such dust aggregates with
$\beta \approx 0.4$ must have survived on the surface against jets due
to ice sublimation, the temperature of ice in the nucleus must be kept
below 145\,K, which is much lower than equilibrium temperature
determined by solar irradiation and thermal emission.  These facts
indicate that 9P/Tempel 1 has a dust mantle 
composed of $20\,\mu$m-sized dust aggregates 
with low
thermal conductivities $\sim 1 \, {\rm erg\, cm}^{-1} \, {\rm
K}^{-1}\,{\rm s}^{-1}$. 
}


\maketitle

\section{Introduction}

Short-period comets are likely born in the outer solar system. They
are originally composed of ices and dust particles. When the comets
approach the sun, volatile components of their nuclei start to
sublime. A jet of volatile gases pushes out dust having a large
surface-to-volume ratio and selectively leaves heavy dust of a small
surface-to-volume ratio \citep[][for a review]{prialnik}. Therefore,
relatively compact particles form the so-called dust mantle of comets.
Dust particles emitted from comets give us important information on
the surfaces of the comets related to their histories. Such a dust
particle feels solar radiation pressure and then its orbit is
determined by the ratio $\beta$ of the radiation pressure to solar
gravity, which depends on the morphology, size, and composition of the
particle. The $\beta$ values of
particles are derived from the the shapes of comet tails and trails 
\citep{finson68, fulle89, ishiguro07,ishiguro08}.  Note that
tail and trail particles of masses $10^{-5}$--$10^{-3}$\,g occupy the largest
end of the size distribution of cometary comae, since smaller
particles are blown away by the radiation pressure after leaving the
comet nuclei.  In-situ measurements of dust in the coma of comet
1P/Halley showed that a typical coma dust particle has a mass of
$10^{-10}$--$10^{-9}$\,g \citep{kolokolova}. Because large dust
particles in tails and trails are most likely a minor component on the cometary
surface, we cannot derive the $\beta$ values of typical dust particles
on surfaces of comets from tail and trail observations. Therefore, the
determination of the $\beta$ value from observations of comet comae is
preferable. 
\citet{hayward} attempted to derive the $\beta$ value from
barred structures formed in a coma of comet Hale-Bopp(C/1995 O1) and
found $\beta < 1$. 
Owing to the fact that a continuous anisotropic ejection of
dust particles in a coma blurs the trajectory of the
particles, it is difficult to tightly constrain the $\beta$
value. 

The advent of a promising occasion for a tight constraint on the $\beta$
value was the Deep Impact (DI) mission, which successfully collided a
366\,kg impactor with Comet 9P/Tempel 1 \citep{ahearn}. The resulting
ejecta plume was observed by space telescopes and by telescopes on the
Earth, which emerged after the DI collision, expanded in about a day,
and then vanished in several days. The plume produced by such a single
collision contained typical dust on the cometary surface and the
dissipation of the plume reflected the blow-away of the dust due to the
radiation pressure. The $\beta$ ratio is estimated to be 0.08--1.9 based
on data analyses of a temporal evolution on the brightness peak position 
of the plume at the
direction to the sun or the shape of the plume at a certain time
\citep{meech,milani,walker,boehnhardt}. These estimates of $\beta$
unfortunately include large uncertainties, because the analyses
were based on limited observational data. 

In this paper, we succeed for the first time in obtaining a tight
constraint on the $\beta$ value of typical dust particles on the surface
of the comet, by taking into account both the temporal evolution in the
distance to the plume leading edge at the solar direction and the plume shape at the
time when the shape is significantly modified by the radiation
pressure. We then use the obtained $\beta$ values to discuss the
physical properties of dust particles, which were originally on the
surface of the nucleus and produced by the DI impact. The successful
determination of the $\beta$ ratio enables us to estimate the mass and
size of the dust particles based on the observational and theoretical
considerations and the temperature of cometary ices, which has been kept
low due to the existence of a dust mantle on the nucleus. We will show
that the thermal conductivity and porosity of the dust mantle could also
be deduced from our estimate of the $\beta$ value.

\section{Model}
\label{sc:model}

Dust particles were ejected from the nucleus of comet Tempel 1 by the DI 
event. These ejecta made an ejecta plume. 
According to an impact cratering theory, 
higher velocity ejacta come from a shallower layer of the target
surface and ejecta from the deeper region have 
lower velocities \citep[e.g.,][]{croft}. The ejecta in the leading edge of the plume 
are thus composed of dust particles on the surface of the comet nucleus,
while the ejecta from the deeper region are mainly in the inner plume. 
Therefore, we could obtain the information of dust particles on the surface of
cometary nucleus from the leading edge of the plume. 

The motion of the ejecta is mainly governed by radiation pressure from the sun. 
The observation from Earth
showed that the ejecta plume expanded isotropically around the impact
direction within several hours from the DI event \citep{feldman07}. We
assume that particles ejected from the surface just after the DI collision have
a velocity projected to the plane perpendicular to the line of sight,
$v_0$.  It should be noted that $v_0$ is lower than the real
ejecta velocity by a factor 1.5--2 \citep{ipatov}. 
We treat their
orbits on the plane. We set the
$x$-axis in the direction to the sun on the plane and the $y$-axis 
perpendicular to the $x$-axis.  The position $(x,y)$ of an ejecta at
time $t$ after the collision of DI impactor is given by
\begin{eqnarray}
x &=& t v_{0} \cos \gamma - t^2 \beta G M_\odot \sin \alpha / 2 D^2,\label{eq:loc_x} \\
y &=& t v_{0} \sin \gamma,\label{eq:loc_y} 
\end{eqnarray}
where $D \approx 1.5$\,AU is the distance of Comet 9P/Tempel 1 from the
sun at the DI event, $\alpha \approx 48^\circ$ is the angle between the
line of sight and the incident radiation from the sun, $G$ is the
gravitational constant, and $\gamma$ is the angle between the direction
of the projected velocity and the $x$ axis.  The $\beta$ ratio is given
by \citep[e.g.,][]{burns}
\begin{equation}
 \beta = \frac{C_{\rm pr} L_\odot}{4 \pi c G M_\odot m},\label{eq:beta} 
\end{equation}
where $C_{\rm pr}$ is the radiation pressure cross section of the
ejecta, $m$ is its mass, $L_\odot$ is the solar luminosity, $M_\odot$ is
the solar mass, and $c$ is the speed of light.

\begin{figure}[htbp]
\includegraphics[width=\linewidth]{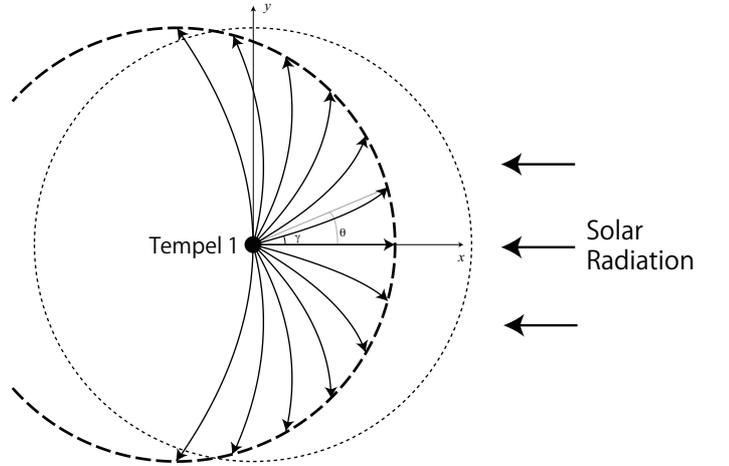}
\caption{
Schematic for the ejecta plume leading edge profile. 
The profile is modified by solar radiation (dashed lines). 
For reference, the profile without the radiation is represented by
the circle with dotted lines. 
\label{fig:schematic}}
\end{figure}

Although the ejecta caused by the DI have various values of $\beta$ and
$v_0$, not all of them are observable. 
\citet{jorda} investigated the ejecta plume profile adopting 
distributions of $\beta$ and $v_0$ and found that the total cross section of dust
particles was almost determined by single values for $v_0$ and $\beta$. 
We therefore assume dust particles have a single value of $\beta$ and eject
with a velocity $v_0$. 
In Fig.~\ref{fig:schematic}, the dashed line given by Equations
(\ref{eq:loc_x}) and (\ref{eq:loc_y}) with certain $\beta$ and $v_0$ 
is hereafter called the leading edge of the plume at a certain time
after the DI collision, where we assume the ejection is
isotropic. 
The plume profile is determined by the $\beta$ and $v_0$
values. 
Since the radiation pressure modifies the plume shape, the
distance of the plume leading edge from Tempel 1 depends on the angle
$\theta = \arctan(y/x)$; the distance at around $\theta \approx 0$ is
shorter than that at a large $|\theta|$. The spatial distribution thus
arrows us to constrain $\beta$ and $v_0$. Nevertheless, since the
observed plume shape includes errors, $\beta$ and $v_0$ could not be
determined solely from the profile at a certain time. The time
evolution of the leading distance at $\theta = 0^\circ$ could give a further
constraint to fix these values.

\section{Results}

\begin{figure}[htbp]
\includegraphics[width=\linewidth]{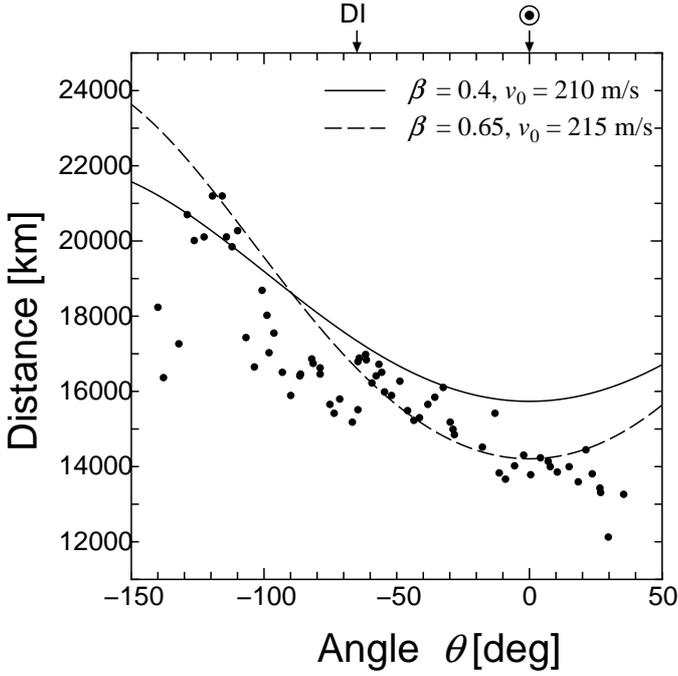}
\caption{
Distance of the
ejecta plume leading edge from Tempel 1 as a function of angle to solar
direction on the view plane, $\theta = \arctan(y/x)$.  Dots represent
observational data at 25 hours after the Deep Impact collision by
\citet{walker} (the brightness contour with 2ADU).
Equations~(\ref{eq:loc_x}) and (\ref{eq:loc_y}) gives the ejecta plume
front taking $\beta = 0.4$ and $v_0 = 210\,{\rm m \, s}^{-1}$ (solid curve) and $\beta
= 0.6$ and $v_0 = 215\,{\rm m\,s}^{-1}$ (dashed curve).  The direction of initial
plume emergence is indicated by DI and that of the sun is by $\odot$.
\label{fig:walker}}
\end{figure}

At first, we focus on the spatial distribution of ejecta plume. 
Hubble Space Telescope observations within several hours after the DI
collision showed that the plume of
ejecta are almost isotropic for the front side of the impact within
$\theta = -145^\circ$--$35^\circ$ \citep{feldman07}; we therefore assume
an isotropic distribution of $\gamma$. The shape of the plume was then
significantly modified after $t \approx 1$\,day
\citep[e.g.,][]{tozzi07,walker}.  \citet{walker} presented the distance
of the plume front from the nucleus at $t = 25$\,hours after the impact.
In Fig.~\ref{fig:walker}, we compare the model calculations (solid and
dashed curves)
given by Equations~(\ref{eq:loc_x}) and (\ref{eq:loc_y}) with the
observational data (dots) by \citet{walker}.  Large $\beta$ leads to the
substantial variation of the distance at the solar direction, while the
variation of $v_0$ produces the difference of the distance independent
of $\theta$.  We have found several possible combinations of $\beta$ and
$v_0$ to explain the data: e.g., $v_0 = 210 \,{\rm m \,s }^{-1}$ and $\beta = 0.4$ or
$\beta = 0.65$ and $v_0 = 215 \, {\rm m \,s}^{-1}$ as shown in Fig.~\ref{fig:walker}.
However, there are still uncertainties of the values of $v_0$ and
$\beta$ determined only from the spatial distribution.  Therefore, we
try to give a further constraint on $\beta$ and $v_0$, by taking into
account the time variation of the plume.

\begin{figure}[htbp]
\includegraphics[width=\linewidth]{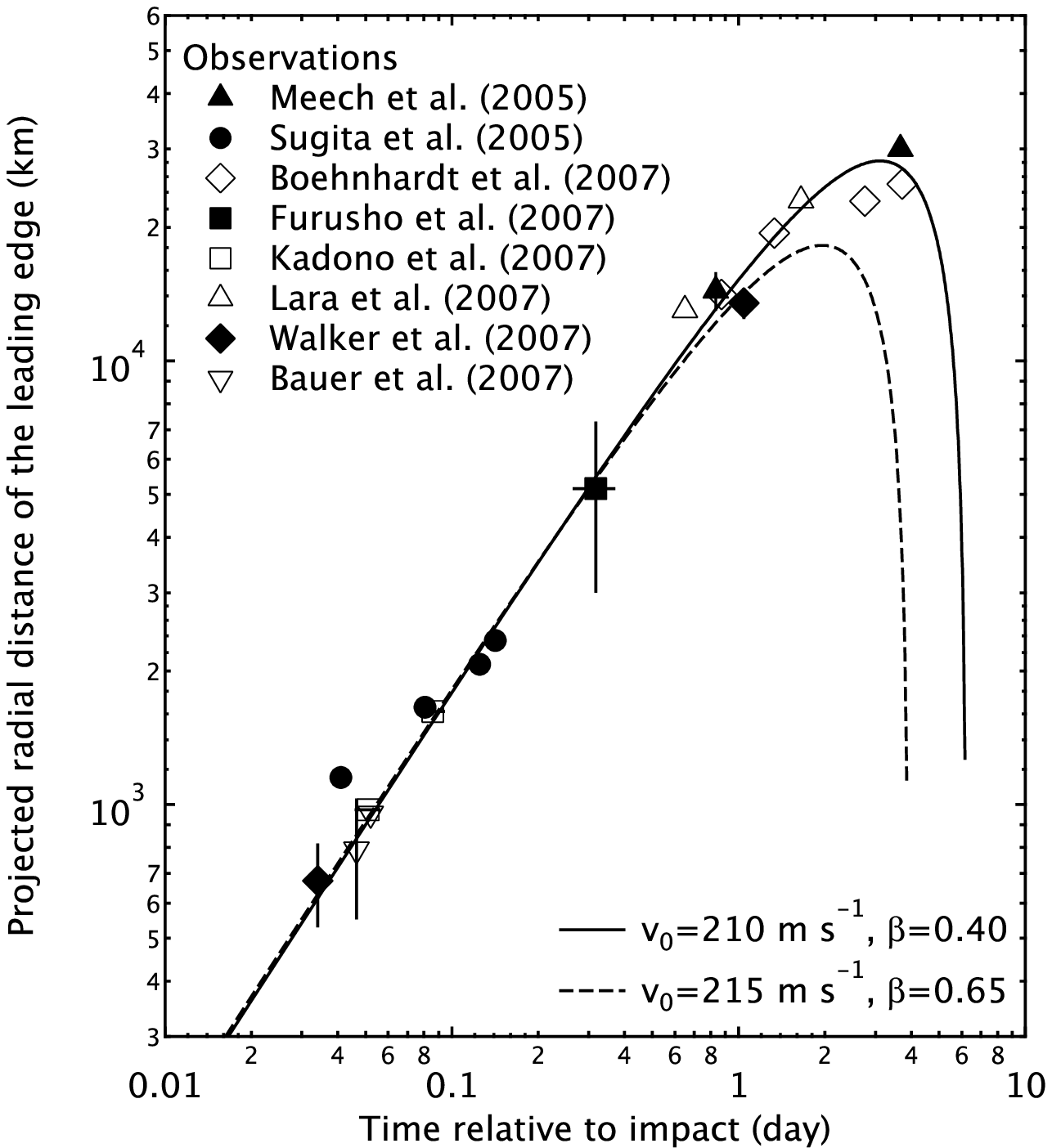}
\caption{
Time evolution of the leading edge distance from observations
 \citep{meech,sugita,boehnhardt,furusho,kadono,lara,walker,bauer}. 
The solid curve is given by 
Equation~(\ref{eq:loc_x}) with  $\beta = 0.4$ and $v_0 = 210 \,{\rm m\,s}^{-1}$. 
The dashed curve is with $\beta = 0.65$ and $v_0 = 215\,{\rm
 m\,s}^{-1}$. 
\label{fig:evolution}
}
\end{figure}

Fig.~\ref{fig:evolution} shows the temporal evolution of the distance 
to the leading edge to the observed ejecta plume at $\theta = 0^\circ$ 
(symbols) in comparison with the model (solid curve). The relation between the
distance and time is linear due to a constant velocity until about one
day, but the ejecta were then significantly decelerated by the radiation
pressure.  The evolution predicted by Equation~(\ref{eq:loc_x}) with
$\beta = 0.4$ and $v_0 = 210 \,{\rm m \, s}^{-1}$ is in good agreement with the distance
evolution obtained from the observations.  The time at the peak
distance, $t_{\rm peak}$, is determined by $d x /dt = 0$, given by
\begin{eqnarray}
  t_{\rm peak} &=& \frac{v_0 D^2 }{\beta G M_\odot \sin
   \alpha}, 
\nonumber
\\
 &\approx& 3 \left(\frac{0.4}{\beta}\right) \left(\frac{v_0}{210\,{\rm
	      m\,s}^{-1}}\right) {\rm days}. 
\end{eqnarray}
The distance decreases by the radiation pressure at $t > t_{\rm peak}$
and the ejecta plume is blown away completely at $t \ga 2 t_{\rm peak}$. If we
choose $\beta = 0.65$ and $v_0 = 215 \,{\rm m} \, {\rm s}^{-1}$, $t_{\rm
peak} \approx 2$ days; the leading distances after 2 days are
inconsistent with the observational data. In addition, 
the maximum projected distance in the sun-ward direction 
obtained from 6 days of observations by \citet{meech} was $3\times 10^4$\,km, which is explained by 
ejecta with $\beta \approx 0.4$ because the peak
distance is given by $v_0 t_{\rm peak}/2 \approx 2.8 \times 10^4 (\beta /0.4)^{-1}
(v_0 / 210\,{\rm m\,s}^{-1})^2$\,km. 
Therefore, taking into account the time
variation and the spatial distribution, $\beta \approx 0.4$ and $v_0
\approx 210 \,{\rm m \, s}^{-1}$ are the likely solutions.

\section{Discussion}

Several authors derived the $\beta$ values within 0.08--1.9 from the
shape of the ejecta plume at a certain time or the temporal evolution of
brightness peak position 
\citep{meech,milani,walker,boehnhardt,richardson,kelley}. 
\citet{schleicher} estimated the lower limit of $\beta$ through a sophisticated simulation using
a Monte Carlo jet model. 
On the other hand, we have applied a simple model described by
Equations~(\ref{eq:loc_x}) and (\ref{eq:loc_y}), whose 
transparent dependences of $\beta$ and $v_0$ easily give a constraint on
them, and have analysed both the leading distance 
evolution of the plume at the
direction to the sun and the shape of the plume at 25\,hours after the DI
event. 
We have found that dust mainly determining the shape of the ejecta plume
has $\beta \approx 0.4$.  Equation (\ref{eq:beta}) gives $C_{\rm pr}/m
\approx 5200\,{\rm cm}^2\,{\rm g}^{-1}$ for $\beta = 0.4$. 
The value corresponds to $0.6\,\micron$ in radius for spherical particles
having internal density $2.5\,{\rm g\,cm}^{-3}$ with $C_{\rm pr}$ given
by geometrical cross section 
\citep{burns}. However, 
dust particles mainly composing the ejecta
plume are likely compact dust aggregates, which account for the
observational data about a silicate-feature strength and a color temperature
\citep{yamamoto}.  According to \citet{yamamoto}, compact aggregates are
assumed to have fractal dimension $D = 2.5$. We apply the description of
fractal aggregates given in \citet{mukai}, using monomers of an aggregate with radius
$0.1\,\micron$ and interior density $2.5\,{\rm g\, cm}^{-3}$.  We
approximate $ C_{\rm pr} \approx A \approx \pi r_{\rm c}^2 $, where $A$
is the projected area of an aggregate and $r_{\rm c} = r_{\rm g}
\sqrt{5/3}$ with $r_{\rm g}$ being the gyration radius of an aggregate.
Dust aggregates with $\beta = 0.4$ approximately have $r_{\rm c} \approx
40\,\micron$ ($m \sim 10^{-8}$\,g ).  On the other hand, based on
in-situ measurements of 1P/Halley, compact dust
aggregates with $r_{\rm c} \approx 10 \, \micron$ ($m \sim
10^{-10}$--$10^{-9}$\,g) contribute most to the cross-section area on
the surface of comet Halley \citep{mcdonnell,kolokolova}. Both DI ejecta from the surface of Tempel 1 and
typical dust aggregates of comet Halley have size of the order of
$10\,\micron$, which may be determined by the temperature history of icy
core of the nuclei.  Therefore, we conclude that the surface layers of
short-period comets are mainly composed of such large aggregates.

A dust particle with the projected area $A$ and mass $m$ on the comet
nucleus with mass $M$ and radius $R$ is blown away due to volatile
sublimation if $A P_{\rm v} \la G M m /R^2$, where $P_{\rm v}$ is the
vapor pressure \citep{prialnik}.  Since $A/m \approx 4 \pi c G M_\odot
\beta / L_\odot$ from Equation~(\ref{eq:beta}) with $C_{\rm pr} \approx
A$, the condition to blown out for particles with $\beta$ is rewritten
as $P_{\rm v} \ga M L_\odot / 4 \pi c R^2 M_\odot \beta$.  Since water ice is
the main volatile material, $P_{\rm v}$ is determined by water ice
sublimation. As shown above, particles ejected by the DI event mainly
have $\beta \approx 0.4$, which survive against ice sublimation on the
nucleus surface. From the fact, we can estimate an upper limit on the
value of vapor pressure, which indicates that the temperature of ice in the comet
nucleus should have been kept lower than about 145\,K.  The ice temperature estimated
above is lower than that on the nucleus surface which is calculated from
the equilibrium between solar irradiation and its thermal emission.
Such a low temperature of ice therefore implies that the nucleus is
covered with a dust layer that has a low thermal conductivity.

Finally we here derive the thermal conductivity of the dust layer. 
The thickness of the dust layer, $L_{\rm d}$, is much smaller than the radius of
the nucleus. 
The heat flux in the dust layer is given by $k_{\rm d} (T_{\rm s}-T_{\rm
i}) / L_{\rm d}
$, where $k_{\rm d}$ is the thermal conductivity of the dust layer,
$T_{\rm s}$ 
is the temperature at the nucleus surface, 
and $T_{\rm i}$ is the
temperature at the boundary between the dust mantle and the ice core. 
The surface temperature obtained form the fitting of a nucleus
spectrum varies from $272\pm7$\,K to $336\pm7$\,K \citep{groussin}. 
The excavation depth
due to the DI impact indicates $L_{\rm d} \sim 100$--$200$\,cm \citep{yamamoto}. 
The conductive flux received on the surface of the ice core should equal
the ice sublimation energy on the ice core surface;
the equilibrium with $T_{\rm i} = 145\,$K results in $k_{\rm d} \approx
1.1 
((T_{\rm s}-T_{\rm i})/190\,{\rm K})^{-1} (L_{\rm d}/200 \, {\rm cm}) \, {\rm erg\, cm}^{-1} \, {\rm K}^{-1}\,{\rm s}^{-1}$. 
Dust particles in the dust mantle of Temple 1 that we have obtained from
$\beta$ have a porosity of 0.99 (or a filling factor $\phi$ of
0.01). 
The mean free path of photons in an aggregate is small enough that the radiative
conductivity is negligible in the interest temperature range. 
From laboratory experiments, 
the thermal conductivity was derived to be $10^{2}$--$10^{3} \, {\rm erg \, cm}^{-1} \, {\rm K}^{-1} \,
{\rm s}^{-1}$ for dust layers with $\phi = 0.1$--0.6 \citep{krause,gundlach}. 
The relation $k_{\rm d} \propto \phi^{p}$ derived from 
theoretical analysis can explain 
experiments for aerogels using power-law index $p \approx 1.5$
\citep{fricke,hrubesh,lu}.  
The thermal conduction for dust layers measured by \citet{krause} is
also well described by 
a power-law formula with $p = 1.5$ (although $p\approx1.8$ is better). 
From the fitting formula, we estimate the thermal conductivity to be $1$--$10 \,
{\rm erg \, cm}^{-1} \, {\rm K}^{-1} \, {\rm s}^{-1}$ for dust
aggregates with $\phi = 0.01$. 
Dust aggregates forming
the dust layer, which can explain several facts obtained from the DI
event, have a high porosity and produce such a low thermal conductivity
of the dust layer.  

\vspace{1cm}

We acknowledge an anonymous reviewer who provided helpful comments to
revise the original manuscript. 
We thank Kazunari Iwasaki, Satoshi Okuzumi, and J\"urgen Blum for valuable discussions. 
This research is supported by the grants from CPS, JSPS (21340040), and
MEXT (23103005).


\end{document}